\documentclass{article}

\usepackage{multicol} 
\usepackage{latexsym}
\usepackage[svgnames]{xcolor} 

\usepackage{times} 

\usepackage{graphicx} 
\graphicspath{{figures/}} 
\usepackage{booktabs} 
\usepackage[font=small,labelfont=bf]{caption} 
\usepackage{amsfonts, amsmath, amsthm, amssymb} 
\usepackage{wrapfig} 

\title{Apollonius  Representation of Qubits}

\author{Oktay K. Pashaev and Tu\u{g}\c{c}e Parlakg\"{o}r\"{u}r \\ \\
Department of Mathematics \\
Izmir Institute of Technology \\
 Izmir, 35430, Turkey}

\begin{document}

\maketitle

\begin{abstract}
We introduce the qubit representation by complex numbers on the set of Apollonius circles with common symmetric points at $0$ and $1$,  related with $|0\rangle$ and $|1\rangle$ states. For one qubit states we find that the Shannon entropy as a measure of randomness is a constant
along Apollonius circles. For two qubit states, the concurence as a characteristic of entanglement is taking constant value for the states on the same Apollonius circle.
Geometrical meaning of concurence as an area and as a distance in the Apollonius representation are found. Then we generalize our results to arbitrary
$n$-qubit Apollonius states and show that the fidelity between given state and the symmetric one, as reflected in an axes, is a constant along Apollonius circles.
For two qubits it coinsides with the concurence. For generic two qubit states we derived Apollonius representation  by three complex parameters
and show that the determinant formula for concurence is related with fidelity for symmetric states by two reflections in a vertical axis and inversion in a circle.
We introduce the complex concurence and an addition formula for Apollonius states and show that for generic two qubit states its modulus  satisfies the law of cosine.
Finally, we show that for two qubit Apollonius state in bipolar coordinates, the complex concurence is decribed by static one soliton solution of the nonlinear Schr\"odinger equation.

\end{abstract}


\let\thefootnote\relax\footnote{Extended version of poster presentation in "Quantum foundations summer school" and "Contextuality
workshop", ETH Zurich, Switzerland, 18-23 June, 2017; "Mathematical Aspects of Quantum
Information", Cargese, France, 4-8 September 2017}


\section{Introduction}
The qubit as a unit of quantum information traditionally is represented by a point on the Bloch sphere. In the coherent state
form it corresponds to to complex number $z$ in extended complex plane $C\bigcup \{\infty\}$, where state $|1\rangle$ is related with $\infty$.
Here we introduce the qubit representation by complex numbers on the set of Apollonius circles with common symmetric points at $0$ and $1$, corresponding to
$|0\rangle$ and $|1\rangle$ states. Apollonius circles are defined as the set of points with given ratio of distances from two fixed points. Pappus d'Alexandrie (290-350 AD) in Book VII of his Collections credits discovery of these circles to Apollonius of Perga (250-170 BC) (Figure 1)
though such circles was considered before by Aristotle (384-322 BC) in Meteorologica \cite{ros}
In recent studies on method of images in hydrodynamics we have solved problem for concentric circles \cite{py}, which can be conformally mapped to arbitrary position of two cylinders in the flow as Apollonius circles. This method of images implies reflection of quantum states in constructing entangled states \cite{pg}, \cite{p14}, and allows us to introduce the Apollonius representation of qubit states.

\begin{center}\vspace{0.25cm}
\includegraphics[width=0.40\linewidth]{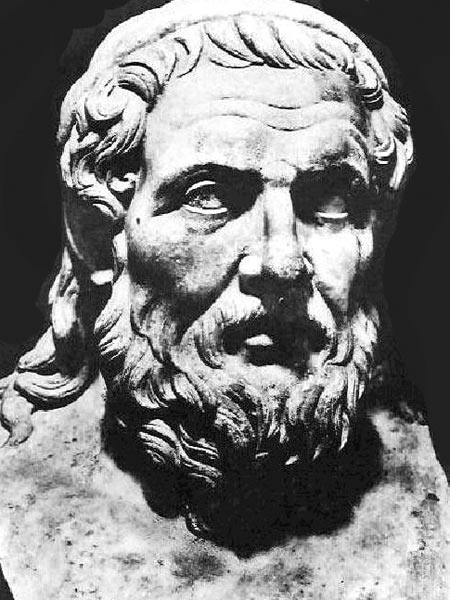}
\captionof{figure}{\color{Navy} Apollonius of Perga (252-170 BC)}
\end{center}\vspace{0.25cm}

\section{One Qubit in Coherent State Representation}
\color{Black} 

One qubit state
\begin{equation}
|\theta, \varphi \rangle = \cos \frac{\theta}{2}|0\rangle + \sin \frac{\theta}{2} e^{i\varphi} |1\rangle
\end{equation}
is determined by a point $(\theta, \varphi)$ on the Bloch sphere. If we project the Bloch sphere to the complex plane, then we
get the qubit coherent state
\begin{equation}
|z\rangle = \frac {|0\rangle + z |1\rangle}{\sqrt {1+ |z|^2}},
\end{equation}
where complex number $z=\tan{\frac{\theta}{2}e^{i \varphi}}$ denotes the stereographic projection (Figure 2).

\begin{center}\vspace{0.25cm}
\includegraphics[width=0.40\linewidth]{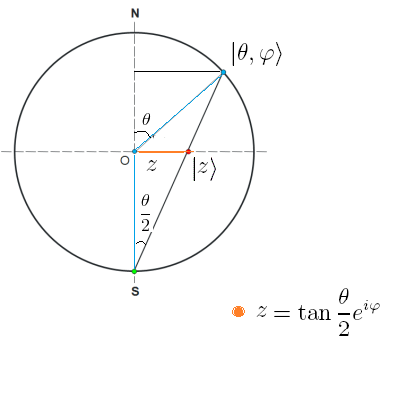}
\captionof{figure}{\color{Navy}Bloch sphere stereographic projection}
\end{center}\vspace{0.25cm}

\begin{center}\vspace{0.25cm}
\includegraphics[width=0.40\linewidth]{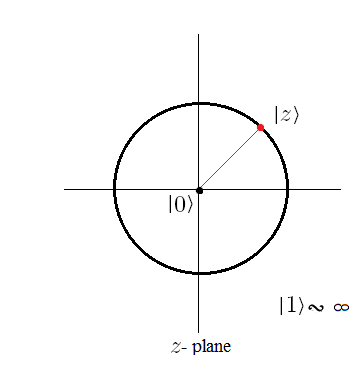}
\captionof{figure}{\color{Navy}Coherent states plane}
\end{center}\vspace{0.25cm}

In this representation $|0 \rangle $ state corresponds to the origin $z = 0$, but the  state $|1\rangle$ is going to infinity, which produces some
disadvantage (Figure 3). However,  by the Hadamard gate we can  move it to a finite point in the plane.
As a result  we get a new representation with state $|0\rangle$ at 1 and state $|1\rangle$ at -1:
\begin{equation}
H |z\rangle = |b\rangle = \frac {(1+z)|0\rangle + (1-z) |1\rangle}{\sqrt{2} \sqrt {1+ |z|^2}}\,.
\end{equation}
To get ordered basis qubits  $|0\rangle$ and $|1\rangle$ at positions $-1$ and $1$ correspondingly, we can apply another circuit diagram:
{\color{Navy}
$$ |z\rangle \line(1,0){50}\fbox{\rule[-.3cm]{0cm}{1cm} Y}  \line(1,0){50}  |-\frac{1}{z}\rangle  \line(1,0){50}   \fbox{\rule[-.3cm]{0cm}{1cm} H}   \line(1,0){50}  |c\rangle                   $$}
where the state (Figure 4)
\begin{equation}
|c\rangle = \frac {(z-1)|0\rangle + (z+1) |1\rangle}{\sqrt{2} \sqrt {1+ |z|^2}}\,.
\end{equation}

\begin{center}\vspace{0.25cm}
\includegraphics[width=0.40\linewidth]{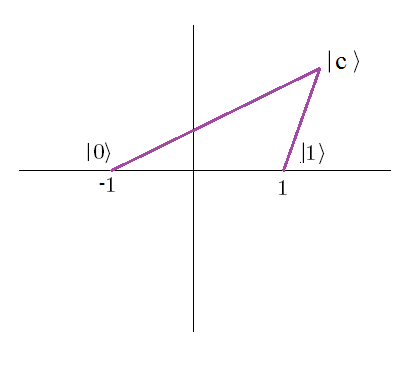}
\captionof{figure}{\color{Navy}Symmetric State}
\end{center}\vspace{0.25cm}

To compare the relation between a qubit and a bit in a more direct way, we like to fix positions of our states $|0\rangle$ and $|1\rangle$ at points $0$ and $1$ correspondingly. Then we replace $z$ to $2z-1$ (scaling and translation) and  get one qubit state $|a\rangle$ in the form
\begin{equation}
|a\rangle = \frac {(z-1)|0\rangle + z |1\rangle}{\sqrt {|z-1|^2 + |z|^2}}\,,
\end{equation}
which we call the Apollonius qubit representation (Figure 5).
\begin{center}\vspace{0.25cm}
\includegraphics[width=0.50\linewidth]{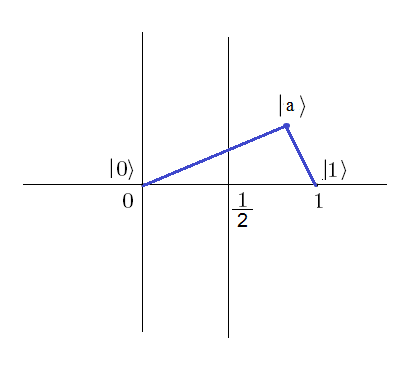}
\captionof{figure}{\color{Navy}State $|a\rangle$}
\end{center}\vspace{0.25cm}
Probabilities to measure state $|0\rangle$ or $|1\rangle$ are:
\begin{equation}
p_0 = \frac {|z-1|^2}{|z-1|^2 + |z|^2}\,,\end{equation}
\begin{equation}
p_1 = \frac {|z|^2}{|z-1|^2 + |z|^2}\,,
\end{equation}
where $p_0+p_1=1$ and the ratio of probabilities is
\begin{equation}
\frac{p_0}{p_1} = \frac{|z|^2}{|z-1|^2} \equiv r^2\,.\label{ratio}
\end{equation}
\textbf{Apollonius Circle Definition}:
A circle can be defined as the set of points in plane that have specified ratio of distances from two fixed points. The ratio is $\frac{|z-a|}{|z-b|}=r$, where $a$ and $b$ are common symmetric points playing role of the fixed points.
In our case $0$ and $1$ are symmetric fixed points, and ratio of probabilities (\ref{ratio}) is  constant along the Apollonius circles (Figure 6).
\begin{center}\vspace{0.25cm}
\includegraphics[width=0.80\linewidth]{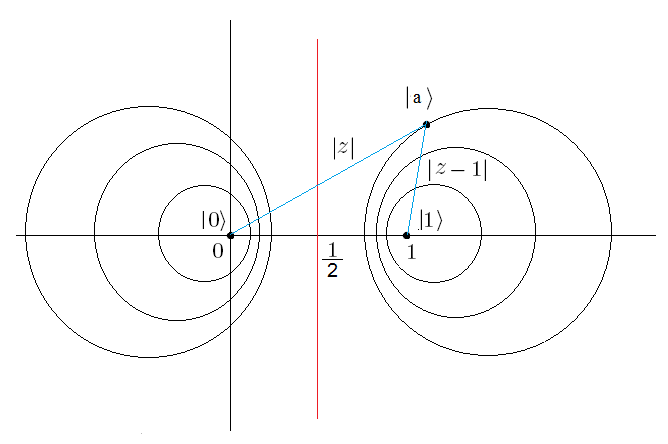}
\captionof{figure}{\color{Navy}Apollonius One Qubit State}
\end{center}\vspace{0.25cm}
The Apollonius circles are determined by $r$ completely, so that the position and the radius of the circle are respectively: $x_0=\frac{r^2}{r^2-1}$ and $r_0=\frac{r}{|r^2-1|}$.\\
\textbf{\textcolor[rgb]{0.00,0.07,1.00}{Apollonius States and Entropy}:}
For Apollonius state $|a\rangle $ we have corresponding probabilities to measure state $|0\rangle$ and $|1\rangle$
\begin{equation}
p_0 = |\langle 0|z\rangle|^2=\frac {|z-1|^2}{|z-1|^2 + |z|^2}= \frac{1}{1+r^2}\,,\end{equation}
\begin{equation}
p_1 = |\langle 1|z\rangle|^2=\frac {|z|^2}{|z-1|^2 + |z|^2} = \frac{r^2}{1+r^2}\,.
\end{equation}
The level of randomness for Apollonius state $|a\rangle$ is determined by the Shannon entropy $$ H =-p_0\log_2p_0 - p_1\log_2p_1$$ and give us following result $$H(r^2)= \log_2(1+r^2)- \frac{r^2}{1+r^2}\log_2r^2.$$ This  means that the Shannon entropy or level of randomness is constant along Apollonius circles (Figure 7). \\
\textbf{\textcolor[rgb]{0.00,0.07,1.00}{Maximally Random States}:}

\begin{center}\vspace{0.25cm}
\includegraphics[width=0.80\linewidth]{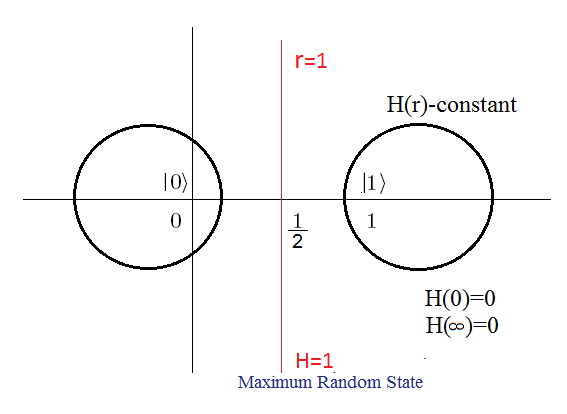}
\captionof{figure}{\color{Navy}Entropy on Apollonius circles}
\end{center}\vspace{0.25cm}

To find maximally random state we take derivative with respect to $r^2$
\begin{equation}
\frac{dH}{dr^2}=-\frac{1}{(1+r^2)^2}\log_2r^2 = 0 \hskip0.5cm \Rightarrow \hskip0.5cm r=1
\end{equation}
The second derivative gives
\begin{equation}
\frac{d^2H}{(dr^2)^2}=-\frac{2}{(1+r^2)^3}\log_2r^2 - \frac{1}{(1+r^2)^2}\frac{1}{r^2\ln2}
\end{equation}
\begin{equation}
H''|_{r=1}=-\frac{1}{4\ln2}<0
\end{equation}

\begin{center}\vspace{0.25cm}
\includegraphics[width=0.50\linewidth]{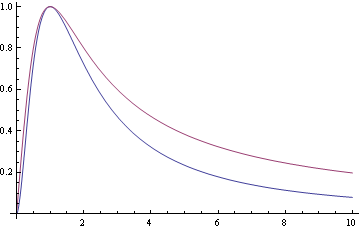}
\captionof{figure}{\color{Navy}Entropy (blue line) and fidelity (pink line) between symmetric states versus $r$}
\end{center}\vspace{0.25cm}

and implies that $r=1$ is the local maximum.
Apollonius circles are level curves of the state randomness (constant entropy $H$ level curves). The maximally random states with $H=1$ are located at vertical line
$\Re z = \frac{1}{2}$. For computational basis states we have zero entropy: $H(0) =0$ and $H(\infty) =0$ (Figure 7).

Another characteristics which is constant along Apollonius circles is the fidelity between symmetric states, reflected in vertical axes
$\Re z = \frac{1}{2}$. This reflection corresponds to substitution $z \rightarrow 1 - \bar z$ and gives the  symmetric state
\begin{equation}
|a_s\rangle = \frac{-\bar z |0\rangle + (1-\bar z)|1\rangle}{\sqrt{|z-1|^2+|z|^2}}\,,\end{equation}
with fidelity
\begin{equation}
F=|\langle a_s |a\rangle| = \frac{2|z||z-1|}{|z-1|^2+|z|^2}.
\end{equation}
In Figure 8 we show the entropy and the fidelity versus $r$. Both curves reach maximal value at $r=1$ and vanish at origin and at infinity.
Comparison of these curves show that maximally random state corresponds to maximal fidelity between symmetric states and it happens when these states belong to the line $\Re z = \frac{1}{2}$.
By increasing the geometrical distance between them we decrease the level of randomness. So that states $|0\rangle$ and $|1\rangle$
as maximally far symmetric states are orthogonal and have fidelity vanishing.

For the distance between symmetric states we get formula
\begin{equation}
\|  |a\rangle - |a_s \rangle         \| = 2 \frac{|\Re z - \frac{1}{2}|}{\sqrt{|z-1|^2+|z|^2}}\,,
\end{equation}
showing that the distance reaches maximal value for orthogonal states at $z=0$ and $z=1$ and on the vertical line $\Re z - \frac{1}{2}$ it vanishes.

We can introduce another distance characteristics in terms of fidelity
\begin{equation}
d = \sqrt{1 - F^2}\,,
\end{equation}
so that for our Apollonius qubit state and the symmetric one we find
\begin{equation}
d = \frac{||z-1|^2-|z|^2|}{|z-1|^2+|z|^2}= \frac{|1-r^2|}{1+r^2} \label{distance}
\end{equation}
This formula shows that distance between symmetric states depends only on Apollonius circle and is determined by its parameter $r$.
It is invariant under substitution $r \rightarrow 1/r$, corresponding to the pair of symmetric circles as reflections in axis $\Re z = \frac{1}{2}$.
For states on the line with $r=1$ we have the minimal distance $d = 0$. For  $r=0$ and $r=\infty$, corresponding to states $|0\rangle$ and
$|1\rangle $ respectively, which are orthogonal states, we find that the distance take maximal value $d = 1$. This value coincides with geometrical distance
between corresponding points $0$ and $1$. As we can see the distance (\ref{distance}) has geometrical meaning of distance between symmetric states
on Aollonius cirles with values $r$ and $1/r$ and is just distance between two points of intersecting Apollonius circles with real line interval
$[0,1]$.

\section{Apollonius Two Qubit States}
\color{Black} 

By applying the $CNOT$ gate to the product state: {\color{Navy}
$$ |a\rangle \otimes|0\rangle \line(1,0){50}\fbox{\rule[-.3cm]{0cm}{1cm} CNOT}    \line(1,0){50}  |A\rangle                   $$}
we get the Apollonius two qubit state (Figure 9)
\begin{equation}
|A\rangle = \frac{(z-1) |00\rangle + z|11\rangle}{\sqrt{|z-1|^2+|z|^2}}\,.\end{equation}

\begin{center}\vspace{0.25cm}
\includegraphics[width=0.80\linewidth]{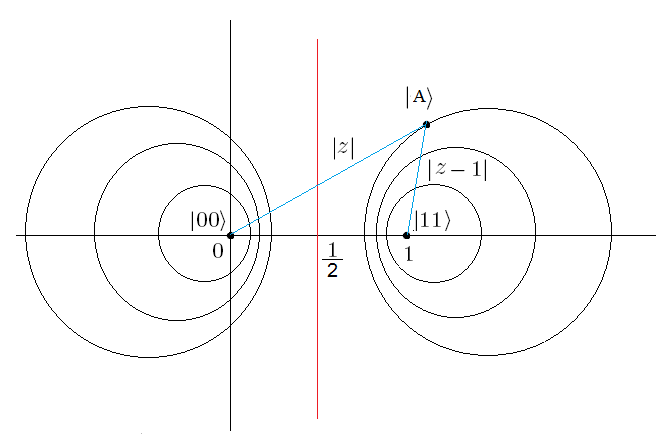}
\captionof{figure}{\color{Navy}Apollonius Two Qubit State}
\end{center}\vspace{0.25cm}

The concurence for this state is (Figure 10)
\begin{equation}
C =\frac{2|z||z-1|}{|z-1|^2+|z|^2}=\frac{2r}{1+r^2}\,,\nonumber
\end{equation}
where $r=\frac{|z|}{|z-1|}$. The concurrence depends on $r$ and $r$ depends on Apollonius circle $\Rightarrow$ concurrence and Apollonius circle are related $\Rightarrow$

\begin{center}\vspace{0.25cm}
\includegraphics[width=0.60\linewidth]{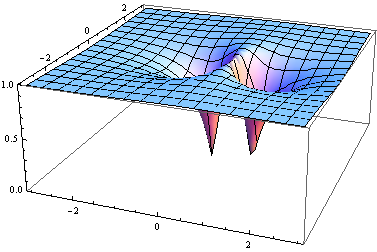}
\captionof{figure}{\color{Navy} Concurence 3d}
\end{center}\vspace{0.25cm}

\begin{center}\vspace{0.25cm}
\includegraphics[width=0.60\linewidth]{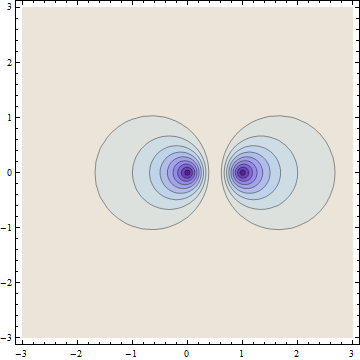}
\captionof{figure}{\color{Navy} Concurence contour plot}
\end{center}\vspace{0.25cm}

\emph{"Concurrence is a constant along Apollonius Circle for given $r$ "} (Figure 11)\\
If $r=1$ $\Rightarrow$ we have the line $Re(z)=\frac{1}{2}$ of quantum states with $C_{max}= 1$, while  states $|00\rangle$ and $|11\rangle$ with $C_{min} = 0$, correspond to common symmetric points for Apollonius circles.\\

\subsection{Geometrical Meaning of Concurrence}
\color{Black} 

For two qubits Apollonius state we have simple geometrical meaning of the concurence. Since concurence is the same for any point on the given
Apollonius circle, we consider intersection of this circle with the orthogonal one $|z-\frac{1}{2}| = \frac{1}{4} $.
In Figure 12a we can see that concurrence is determined as double area of the rectangle and in Figure 12b as a distance between intersection points.

\begin{center}\vspace{0.25cm}
\includegraphics[width=1.0\linewidth]{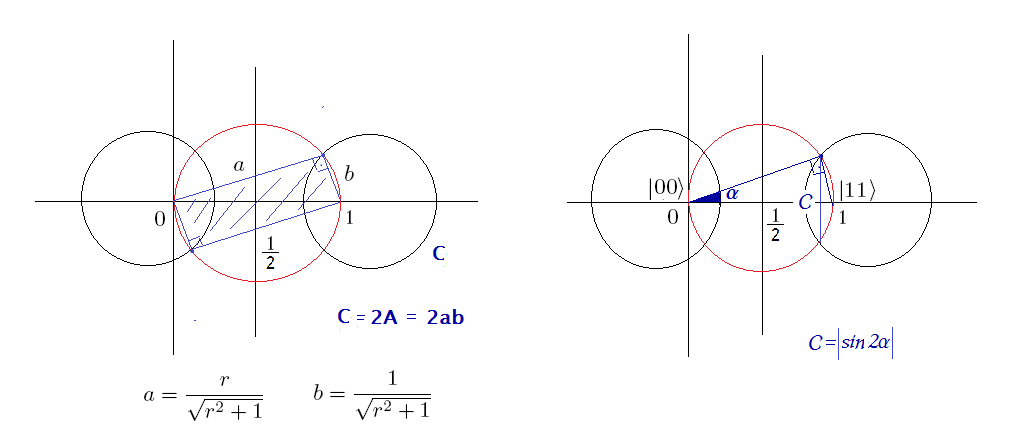}
\captionof{figure}{\color{Navy}a) Concurrence as an area, b) concurence as a distance}
\end{center}\vspace{0.25cm}

\subsection{Concurrence and Reflection Principle}
\color{Black} 

For Apollonius two qubit state $|A\rangle$ we take reflection with respect to the line $Re(z)=\frac{1}{2}$, giving the symmetric two qubit state (Figure 13)

\begin{equation}
|A_s\rangle = \frac{-\bar z |00\rangle + (1-\bar z)|11\rangle}{\sqrt{|z-1|^2+|z|^2}}\,,\end{equation}

\begin{center}\vspace{0.25cm}
\includegraphics[width=0.4\linewidth]{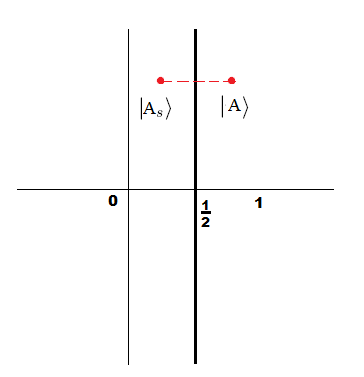}
\captionof{figure}{\color{Navy}Symmetric qubit states}
\end{center}\vspace{0.25cm}
and fidelity between symmetric states is just the concurence
\begin{equation}
F=|\langle A_s |A\rangle| = \frac{2|z||z-1|}{|z-1|^2+|z|^2} = C\,.
\end{equation}

\subsection{Generalization to n-qubit Apollonius states}
\color{Black} By circuit :

{\color{Navy}
$$ |a\rangle \otimes|0\rangle ...|0\rangle \otimes|0\rangle\line(1,0){20}\fbox{\rule[-.3cm]{0cm}{1cm} CNOT}\otimes... I\otimes  I \line(1,0){20}...
 \line(1,0){20}I\otimes I ... \otimes\fbox{\rule[-.3cm]{0cm}{1cm} CNOT}
 \line(1,0){20}  |A\rangle                   $$}
 we can generate the $n$-qubit Apollonius state
\begin{equation}
|A\rangle = \frac{(z-1) |00...0\rangle + z|11...1\rangle}{\sqrt{|z-1|^2+|z|^2}}\,.\end{equation}

The corresponding symmetric state is
\begin{equation}
|A_s\rangle = \frac{-\bar z |00...0\rangle + (1-\bar z)|11...1\rangle}{\sqrt{|z-1|^2+|z|^2}}
\end{equation}
and fidelity between these states
\begin{equation}
F=|\langle A_s |A\rangle| = \frac{2|z||z-1|}{|z-1|^2+|z|^2} = \frac{2r}{1+r^2}
\end{equation}
is constant on Apollonius circle with fixed $r$.

\section{Qubit in Bipolar Coordinates}

The Apollonius circle representation suggests that the bipolar coordinates could be useful in description of qubit.
These coordinates have applications in navigation problems and electro-magnetic theory, determining the electric and magnetic field of two infinitely long parallel
cylindrical conductors. For given complex $z = x+ iy$ we introduce
two real variables, $\tau$ and $\sigma$ (Figure 14)
\begin{equation}
z = \frac{e^\tau}{e^\tau - e^{i\sigma}}\,,
\end{equation}
where
\begin{equation}
\frac{|z|}{|z-1|} = r = e^\tau \,.
\end{equation}

\begin{center}\vspace{0.25cm}
\includegraphics[width=0.8\linewidth]{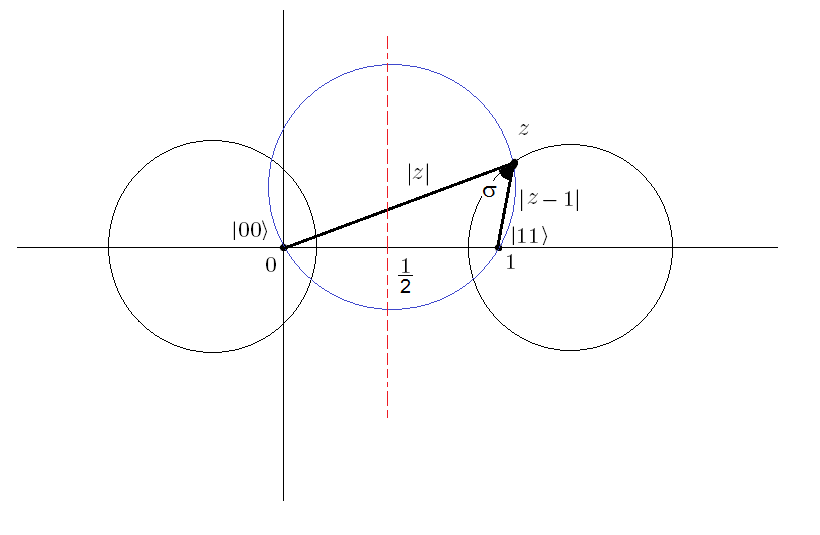}
\captionof{figure}{\color{Navy}Bipolar coordinates}
\end{center}\vspace{0.25cm}

For Cartesian coordinates we have
\begin{eqnarray}
x& =& \frac{1}{2} + \frac{1}{2}\frac{\sinh \tau}{\cosh \tau - \cos \sigma} \,,\\
y& =& \frac{1}{2}\frac{\sin \sigma}{\cosh \tau - \cos \sigma}\,,
\end{eqnarray}
so that
\begin{equation}
z = x + iy = \frac{1}{2} + \frac{1}{2}\frac{\sinh \tau + i \sin \sigma}{\cosh \tau - \cos \sigma}\,.
\end{equation}
 Then for one qubit in bipolar coordinates we have
 \begin{equation}
 |A\rangle = \frac{1}{2}\frac{(e^{i\sigma} - e^{-\tau})|0\rangle + (e^{\tau}- e^{-i\sigma}) |1\rangle}{\sqrt{\cosh \tau (\cosh \tau - \cos \sigma)}}\,.
 \end{equation}
This state up to global phase can be rewritten as
\begin{equation}
|\tau, \sigma \rangle = \frac{e^{i\sigma}|0\rangle + e^\tau |1\rangle}{\sqrt{1 + e^{2\tau}}}.
\end{equation}
For Apollonius two qubit state in bipolar coordinates ina similar way we have
\begin{equation}
|\tau, \sigma \rangle = \frac{e^{i\sigma}|00\rangle + e^\tau |11\rangle}{\sqrt{1 + e^{2\tau}}}\,.
\end{equation}
Applying the determinant formula for concurence of this two qubit state, we find simple expression
\begin{equation}
C = \frac{1}{\cosh \tau} = sech\, \tau\,.\label{concpolar}
\end{equation}
It shows that concurence is not depending on angle $\sigma$ and is constant along the Apollonius circle with given coordinate $\tau$.
This formula suggests to consider the transition amplitude between symmetric states in bipolar coordinates.
For $n$- qubit state it gives the complex fidelity
\begin{equation}
{\cal F} = \langle A_s |A\rangle = F e^{-i\sigma} = \frac{e^{-i\sigma}}{\cosh \tau}\,,
\end{equation}
 which in the case of two qubit states describe the complex version of the concurence
 \begin{equation}
{\cal C} = \langle A_s |A\rangle = C e^{-i\sigma} = \frac{e^{-i\sigma}}{\cosh \tau}\,,\label{cc}
\end{equation}
 such that the modulus of this complex concurence is just the usual concurence (\ref{concpolar}) (Figure 15):
 \begin{equation}
|{\cal C}| = | \langle A_s |A\rangle| = C  = \frac{1}{\cosh \tau}\,.
\end{equation}

\begin{center}\vspace{0.25cm}
\includegraphics[width=0.4\linewidth]{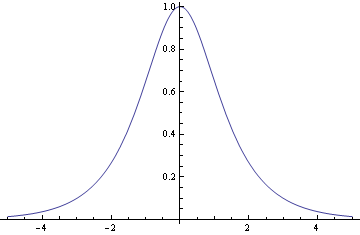}
\captionof{figure}{\color{Navy}Soliton shape for concurence}
\end{center}\vspace{0.25cm}

 It is interesting to notice that complex concurence (\ref{cc}) is the stationary one soliton solution of the Nonlinear
 Schr\"odinger equation
 \begin{equation}
 i {\cal C}_\sigma = {\cal C}_{\tau\tau} + 2 |{\cal C}|^2 {\cal C}.
 \end{equation}
This equation is a nonlinear integrable system admitting arbitrary N-soliton solutions and it has appear in many physical applications from plasma physics to fluid mechanics. We don't know if the above result,
that complex concurence, as a transition amplitude between symmetric Apollonius states, represents soliton of NLS equation has deep meaning,
but existence of such relation is really amazing.

\section{Apollonius representation of generic two qubit state}
The Apollonius states as we have introduced above are characterized by one complex parameter. For one qubit case it is
the generic state. However, for multiple generic qubit states we need to introduce more parameters.
Here we derive Apollonius representation for the generic two qubit state
\begin{equation}
|\psi \rangle = c_{00} | 00 \rangle + c_{01} | 01 \rangle +c_{10} | 10 \rangle +c_{11} | 11 \rangle\,,
\end{equation}
where
\begin{equation}
|c_{00}|^2 + |c_{01}|^2 + |c_{10}|^2 + |c_{11}|^2 =1\,.
\end{equation}
Instead of four complex variables $c_{ij}$, $i,j = 0,1$, we can introduce another set of four complex variables
$\eta$, $\zeta$, $a$ and $b$ according to formulas
\begin{eqnarray}
c_{00}= (\eta -1)\, a,\,\,\,\,\,\,c_{11} = \eta\, a \,,\\
c_{01} = (\zeta -1)\, b,\,\,\,\,\,\,c_{10} = \zeta\, b \,,
\end{eqnarray}
where complex variables $a$ and $b$ we express in terms of complex $\alpha$ and $\beta$ as:
\begin{equation}
a = \frac{\alpha}{\sqrt{|\eta -1|^2 + |\eta|^2}},\,\,\,\,b = \frac{\beta}{\sqrt{|\zeta -1|^2 + |\zeta|^2}}\,.
\end{equation}
Introducing Apollonius two qubit  states
\begin{equation}
|\eta \rangle = \frac {(\eta-1)|00\rangle + \eta |11\rangle}{\sqrt {|\eta-1|^2 + |\eta|^2}},\,\,\,\,
|\zeta \rangle = \frac {(\zeta-1)|01\rangle + \zeta |10\rangle}{\sqrt {|\zeta-1|^2 + |\zeta|^2}}\,,
 \end{equation}
 for the generic state we get superposition
 \begin{equation}
 |\psi \rangle = \alpha |\eta \rangle + \beta |\zeta \rangle\,.
 \end{equation}
Parameters $\alpha$ and $\beta$ can be fixed by normalization condition. We notice that Apollonius states
$|\eta \rangle$ and $|\eta \rangle$ are orthogonal and normalized:
\begin{equation}
\langle \eta | \eta \rangle = \langle \zeta | \zeta \rangle = 1, \,\,\,\,\,\,\langle \eta | \zeta \rangle = \langle \zeta | \eta \rangle = 0.
\end{equation}
It implies $|\alpha|^2 + |\beta|^2 = 1$ and we can choose
\begin{equation}
\alpha = (\xi - 1)\, \lambda,\,\,\,\,\,\,\beta = \xi \, \lambda\,,
\end{equation}
where $\xi$ is an arbitrary complex number and
\begin{equation}
|\lambda| = \frac{1}{\sqrt{|\xi -1|^2 + |\xi|^2}}\,.
\end{equation}
Then by neglecting arbitrary global phase factor we have normalized generic two qubit state in Apollonius representation,
characterized by three arbitrary complex numbers $\eta$, $\zeta$ and $\xi$ :
\begin{equation}
|\psi \rangle = \frac{(\xi -1) |\eta \rangle + \xi |\zeta \rangle}{\sqrt{|\xi -1|^2 + |\xi|^2}}\,.
\end{equation}
Concurence of this state, calculated by the determinant formula is
\begin{equation}
C = \frac{2}{\sqrt{|\xi -1|^2 + |\xi|^2}} \left|   (\xi -1)^2\frac{\eta(\eta -1)}{\sqrt {|\eta-1|^2 + |\eta|^2}}
 - \xi^2 \frac{\zeta(\zeta -1)}{\sqrt {|\zeta-1|^2 + |\zeta|^2}}                  \right|\label{C}
\end{equation}
In particular cases it reduces to previous results
\begin{eqnarray}
\xi = 0 \Rightarrow C = \frac{2 |\eta||\eta -1|}{\sqrt {|\eta-1|^2 + |\eta|^2}} \,,\\
\xi = 1 \Rightarrow C = \frac{2 |\zeta||\zeta -1|}{\sqrt {|\zeta-1|^2 + |\zeta|^2}}\,.
\end{eqnarray}

\subsection{Reflected qubit and concurence}

The concurence formula (\ref{C}) can be derived from the reflection principle for Apollonius generic two qubit state as
\begin{equation}
C = |\langle \psi_s | \psi \rangle |\,,
\end{equation}
where the symmetric qubit state $|\psi_s\rangle $ is coming from reflection of input qubits in three steps.

1) Reflection in complex plane $\eta$ in the vertical line $\Re \eta = \frac{1}{2}$ (Figure 16):

$\eta_s \equiv \eta^* = 1-\bar \eta$

\begin{center}\vspace{0.25cm}
\includegraphics[width=0.4\linewidth]{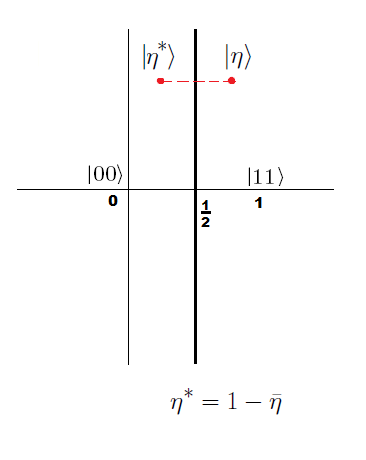}
\captionof{figure}{\color{Navy}Symmetric qubit $| \eta \rangle$}
\end{center}\vspace{0.25cm}

2) Reflection in complex plane $\zeta$ in the vertical line $\Re \zeta = \frac{1}{2}$ (Figure 17):

$\zeta_s \equiv \zeta^* = 1-\bar \zeta$

\begin{center}\vspace{0.25cm}
\includegraphics[width=0.4\linewidth]{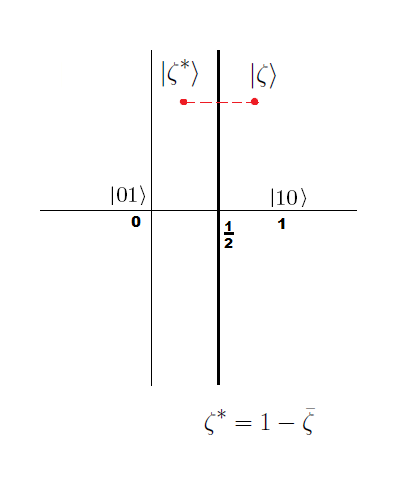}
\captionof{figure}{\color{Navy}Symmetric qubit $| \zeta \rangle$}
\end{center}\vspace{0.25cm}

3) Inversion in complex plane $\xi$ in  circle $|\xi - \frac{1}{2}| = \frac{1}{4}$ Figure (18):

$\xi_s \equiv \xi^* = \frac{1}{2} + \frac{1/4}{\bar \xi - 1/2}$

\begin{center}\vspace{0.25cm}
\includegraphics[width=0.5\linewidth]{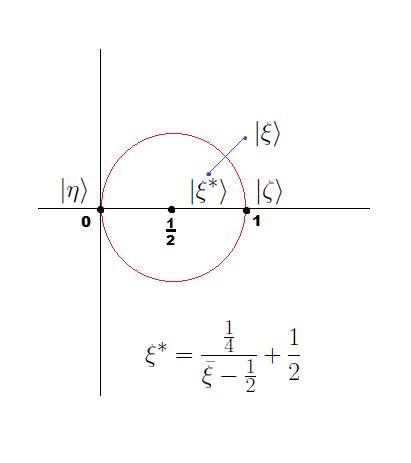}
\captionof{figure}{\color{Navy}Symmetric qubit by inversion $| \xi \rangle$}
\end{center}\vspace{0.25cm}

The resulting state is
\begin{equation}
|\psi_s \rangle = \frac{(\xi^* -1) |\eta^* \rangle + \xi^* |\zeta^* \rangle}{\sqrt{|\xi^* -1|^2 + |\xi^*|^2}}\,,
\end{equation}
or up to global phase
\begin{equation}
|\psi_s \rangle = \frac{(\bar \xi -1) |\eta^* \rangle - \bar\xi |\zeta^* \rangle}{\sqrt{|\xi -1|^2 + |\xi|^2}}\,,\label{psis}
\end{equation}
where symmetric qubit states are
\begin{equation}
|\eta^* \rangle = -\frac{\bar \eta |00 \rangle + (\bar\eta -1) |11 \rangle}{\sqrt{|\eta -1|^2 + |\eta|^2}}\,,
\end{equation}
\begin{equation}
|\zeta^* \rangle = -\frac{\bar \zeta |01 \rangle + (\bar\zeta -1) |10 \rangle}{\sqrt{|\zeta -1|^2 + |\zeta|^2}}\,.
\end{equation}
Calculating the concurence $C = |\langle \psi_s | \psi \rangle |$ we obtain the same result as by determinant formula (\ref{C}).

It is instructive to see how the phase flipping gate action \cite{w} is related with reflection of Apollonius qubits.
Applying the gate
to anti-unitary transformed states
\begin{equation}
K |  \eta \rangle = | \bar \eta \rangle,\,\,\,\,K |  \zeta \rangle = | \bar \zeta \rangle\,,
\end{equation}
we have reflected states
\begin{equation}
Y \otimes Y | \bar \eta \rangle = |\eta^* \rangle \,,
\end{equation}
\begin{equation}
Y \otimes Y | \bar \zeta \rangle = -|\zeta^* \rangle \,,
\end{equation}
and
\begin{equation}
Y \otimes Y | \bar \psi \rangle = \frac{(\bar \xi -1)Y \otimes Y |\\ \bar \eta \rangle + \bar \xi \,Y \otimes Y|\bar \zeta \rangle}{\sqrt{|\xi -1|^2 + |\xi|^2}} =
\frac{(\bar \xi -1) |\eta^* \rangle - \bar\xi |\zeta^* \rangle}{\sqrt{|\xi -1|^2 + |\xi|^2}} = |\psi_s \rangle\,.
\end{equation}

\subsection{Law of Cosines for concurence}

Transition amplitude written in the form
\begin{equation}
\langle \psi_s | \psi \rangle = \frac{(\xi -1)^2  \langle \frac{1}{\bar \eta} | \eta \rangle   - \xi^2 \langle \frac{1}{\bar \zeta} | \zeta \rangle   }{|\xi -1|^2 + |\xi|^2}
\end{equation}
has interesting geometrical interpretation. If we introduce the total complex concurence, ${\cal C} = \langle \psi_s | \psi \rangle$ , and
two partial complex concurences
${\cal C}_{\eta} = \langle \frac{1}{\bar \eta} | \eta \rangle$ and ${\cal C}_{\zeta} = \langle \frac{1}{\bar \zeta} | \zeta \rangle$
then this relation reads as a superposition principle for complex concurences
\begin{equation}
{\cal C} = \mu {\cal C}_{\eta} + \nu {\cal C}_{\zeta}\,,\label{concurence}
\end{equation}
where complex numbers are defined as
\begin{equation}
\mu = \frac{(\xi-1)^2}{|\xi -1|^2 + |\xi|^2}, \,\,\,\,\,\nu = -\frac{\xi^2}{|\xi -1|^2 + |\xi|^2}\,,
\end{equation}
and satisfy
\begin{equation}
|\mu| + |\nu| =1, \,\,\,\,\frac{|\nu|}{|\mu|} = \frac{|\xi|^2}{|\xi -1|^2}  = R^2\,.
\end{equation}
For partial complex concurences we have
\begin{equation}
{\cal C}_{\eta} = \frac{2\eta(\eta -1)}{\sqrt {|\eta-1|^2 + |\eta|^2}}\,,
\end{equation}
\begin{equation}
{\cal C}_{\zeta} = \frac{2\zeta(\zeta -1)}{\sqrt {|\zeta-1|^2 + |\zeta|^2}}\,.
\end{equation}
From (\ref{concurence}) we have the Law of Cosine
\begin{equation}
C^2 = (|\mu| C_\eta)^2 + (|\nu| C_\zeta)^2 - 2 (|\mu| C_\eta)(|\nu| C_\zeta) \cos \Phi
\end{equation}
\begin{center}\vspace{0.25cm}
\includegraphics[width=0.7\linewidth]{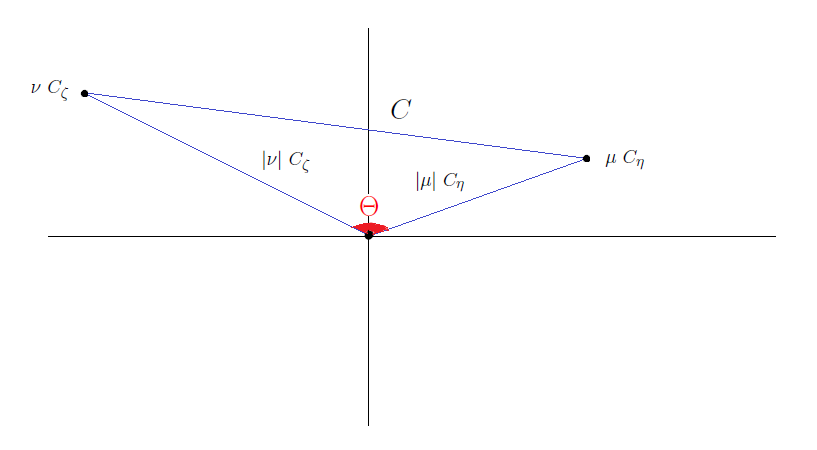}
\captionof{figure}{\color{Navy}Law of Cosine for complex concurence}
\end{center}\vspace{0.25cm}

 for concurences: $C = |{\cal C}|$, $C_\eta = |{\cal C}_\eta|$, and $C_\zeta = |{\cal C}_\zeta|$ (Figure 19).

\color{Black} 




\section{Conclusions}
For arbitrary n-qubit Apollonius states, fidelity between symmetric states is constant along Apollonius circles. In one qubit case it is related with Shannon entropy
and for two qubit states it coincides with concurence. For generic two qubit states we derived Apollonius representation  by three complex parameters
and show that the determinant formula for concurence is related with fidelity for symmetric states by two reflections in a vertical axis and inversion in a circle.
By introduction of complex concurence for generic two qubit states we derived the law of cosine.
For two qubit Apollonius state in bipolar coordinates, the complex concurence is decribed by static one soliton solution of the nonlinear Schr\"odinger equation.
\color{Black} 

\color{Black} 
\subsection*{Acknowledgements} This work is supported by TUBITAK grant 116F206.


\end{document}